\documentclass[aps,prx,twocolumn,showkeys,superscriptaddress]{revtex4-1}

\usepackage{amsmath}
\usepackage{amssymb}
\usepackage{color}
\usepackage{graphicx}
\usepackage{epstopdf}
\usepackage{pxfonts}

\DeclareMathOperator{\arsinh}{arsinh}
\DeclareMathOperator{\artanh}{artanh}

\begin{document}

\title{Surface and zeta potentials of charged permeable nanocoatings }

\author{Elena F. Silkina}
\affiliation{Frumkin Institute of Physical Chemistry and Electrochemistry, Russian Academy of Sciences, 31 Leninsky Prospect, 119071 Moscow, Russia}

\author{Naren Bag}
\affiliation{DWI - Leibniz Institute for Interactive Materials,  Forckenbeckstr. 50, 52056 Aachen, Germany}

\author{Olga I. Vinogradova}
\email[Corresponding author: ]{oivinograd@yahoo.com}

\affiliation{Frumkin Institute of Physical Chemistry and Electrochemistry, Russian Academy of Sciences, 31 Leninsky Prospect, 119071 Moscow, Russia}
\affiliation{DWI - Leibniz Institute for Interactive Materials,  Forckenbeckstr. 50, 52056 Aachen, Germany}

\date{\today }

\begin{abstract}
An electrokinetic (zeta) potential of charged permeable porous films on solid supports  generally exceeds their surface potential, which often builds up to a quite high value itself. Recent work provided a quantitative understanding of zeta potentials of thick,  compared to the extension of an inner  electrostatic diffuse layer, porous films. Here, we consider porous coatings of a thickness comparable or smaller than that of the inner diffuse layer. Our theory, which is valid even when electrostatic potentials become  quite high and accounts for a finite hydrodynamic permeability of the porous materials, provides a framework for interpreting the difference between values of surface and zeta potentials in various situations.  Analytic approximations for the zeta potential in the experimentally relevant limits  provide a simple explanation of transitions between different regimes of electro-osmotic flows, and also suggest strategies for its tuning in microfluidic applications.

\end{abstract}


\maketitle

\renewcommand{\vec}[1]{\boldsymbol{#1}}

\affiliation{DWI - Leibniz Institute for Interactive Materials,
Forckenbeckstr. 50, 52056 Aachen, Germany}

\affiliation{Frumkin Institute of Physical Chemistry and Electrochemistry,
Russian Academy of Sciences, 31 Leninsky Prospect, 119071 Moscow, Russia}

\affiliation{Frumkin Institute of Physical Chemistry and Electrochemistry,
Russian Academy of Sciences, 31 Leninsky Prospect, 119071 Moscow, Russia} %
\affiliation{Lomonosov Moscow State University, 119991 Moscow, Russia}

\affiliation{DWI - Leibniz Institute for Interactive Materials,
Forckenbeckstr. 50, 52056 Aachen, Germany}
\affiliation{Frumkin Institute of
Physical Chemistry and Electrochemistry, Russian Academy of Sciences, 31
Leninsky Prospect, 119071 Moscow, Russia}
\affiliation{Lomonosov Moscow
State University, 119991 Moscow, Russia}



\section{Introduction}\label{sec:introduction}

A century ago \citet{smoluchowski.m:1921} proposed an  equation to describe a plug electro-osmotic flow in a bulk electrolyte that emerges when an electric field $E$ is applied at tangent to a charged solid surface. He related the velocity in the bulk $V_{\infty}$ to the electrokinetic (zeta) potential of the surface $Z$. For canonical  solid surfaces with \emph{no-slip} hydrodynamic  boundary condition, simple  arguments lead to $Z = \Psi_s$, where $\Psi_s$ is the surface (electrostatic) potential. However, the problem is not that simple and has been revisited in last decades. For example, even ideal solids, which are smooth, impermeable, and chemically homogeneous, can modify the hydrodynamic boundary conditions when poorly wetted~\cite{vinogradova.oi:1999}, and the emerging hydrophobic slippage can augment $Z$ compared to the surface potential~\cite{muller.vm:1986,joly.l:2004,maduar.sr:2015,silkina.ef:2019}. Furthermore, most solids are not ideal but
rough and heterogeneous. This can further change, and quite dramatically, the
boundary conditions~\cite{vinogradova.oi:2011} leading to a very rich electro-osmotic behavior and, in some situations (e.g. superhydrophobic surfaces), providing a huge flow enhancement compared to predicted by the Smoluchowski model~\cite{bahga.ss:2009,squires.tm:2008,belyaev.av:2011a}.

The defects or pores of the wettable solids also modify the hydrodynamic boundary condition~\cite{beavers.gs:1967}. Besides, the local electro-neutrality is broken not only in the outer diffuse layer as it occurs for impenetrable surfaces~\cite{anderson.jl:1989,bocquet.l:2010}, but also in the inner one~\cite{ohshima.h:1985}. Moreover, even when the porous coating is electrostatically thick, i.e. includes a globally electro-neutral region, only mobile absorbed ions can react to an applied electric field~\cite{vinogradova.oi:2020}. Consequently, the electric volume force that drives the electro-osmotic flow in the electro-neutral bulk electrolyte is now generated inside the porous material too. This suggests that one can significantly impact the electro-kinetic  response of the whole macroscopic system, i.e. of the bulk electrolyte, just by using various permeable nanometric coatings at the solid support, such as polyelectrolyte networks, multilayers, and brushes~\cite{cohenstuart.m:2010,chollet.b:2016,ballauff.m:2006,vinogradova.oi:2006,das.s:2015}, or  ultrathin porous membrane films~\cite{vandenBerg.a:2007,pina.mp:2011,lukatskaya.m:2016}.

The emerging flow is strongly coupled to the  electrostatic potential profile that sets up self-consistently, so the latter becomes a very important consideration  in electroosmosis involving porous surfaces. Electrostatic potentials, $\Psi_s$ and $\Psi_0$ at the solid support, have been studied theoretically over several decades.
 In most of these studies weakly charged surfaces or thick compared to their inner screening length porous films have been considered~\cite{donath.e:1979,ohshima.h:1985,ohshima.h:1990a,ohshima.h:1995}. Very recently \citet{silkina.ef:2020b} reported a closed-form analytic solution for $\Psi_0$, obtained without a small potential assumption, which is valid for porous films of any thickness. These authors also proposed a general relationship between $\Psi_s$ and $\Psi_0$, but made no attempts to derive simple asymptotic approximations for surface potentials that could be handled easily.

The connection between the electro-osmotic velocity and electrostatic potentials have been reported by several groups~\cite{donath.e:1979,ohshima.h:1990a,ohshima.h:1995,duval.jfl:2004,duval.jfl:2005,ohshima.h:2006}, and these models are frequently invoked in the interpretation of the electrokinetic data~\cite{barbati.ac:2012}. However, despite its fundamental and practical significance, the zeta-potential of porous surfaces has received so far little attention,
and its relation to $\Psi_s$ has remained  obscure until recently. Some authors concluded that the zeta-potential `loses its significance'~\cite{ohshima.h:1990a}, `irrelevant as a concept'~\cite{yezek.lp:2004} or `is undefined and thus nonapplicable'~\cite{duval.jfl:2004}, while others reported that $Z$ typically exceeds $\Psi_s$~\cite{sobolev.vd:2017,chen.g:2015},  but did not attempt to relate their results to the inner flow and emerging liquid velocity at the porous surface. This was taken up only recently in the paper by \citet{vinogradova.oi:2020}, who carried out calculations of the zeta potential for thick coatings of both an arbitrary volume charge density and  a finite hydrodynamic permeability. These authors  predicted that $Z$ is generally augmented compared to the surface electrostatic potential, thanks to a liquid \emph{slip} at their surface emerging due to an electro-osmotic flow in the enriched by counter-ions porous films.  However, this work cannot be trivially extended to the case of non-thick films, where inner electrostatic potential profiles are always, and often essentially, inhomogeneous. These profiles can be calculated assuming that electrostatic potentials are low~\cite{ohshima.h:1985}, but such an assumption becomes unrealistic in many situations. Recently, \citet{silkina.ef:2020b} derived rigorous upper and lower bounds on $Z$ of non-thick films, by lifting an assumption of low electrostatic  potential. However, we are unaware of any prior work that investigated the connection of the zeta potential of non-thick films with their finite  hydrodynamic permeability.

In this paper, we provide analytical solutions to electro-osmotic flows in and outside uniformly charged non-thick porous coatings, with the focus on their zeta potential and its relation to the surface potential. Ionic solutions are described using the non-linear mean-field Poisson-Boltzmann theory. For simplicity, here we treat only the symmetric monovalent electrolyte, but it is rather straightforward to extend our results to multivalent ionic systems. As any approximation, the Poisson-Boltzmann formalism has its limits of validity, but it always describes very accurately the ionic distributions for monovalent ions in the typical concentration range from 10$^{-6}$ to 10$^{-1}$ mol/L~\cite{andelman.d:2006book}. Since in this concentration range $\lambda_D$ decreases from ca. 300 down to 1 nm~\cite{israelachvili.jn:2011}, the non-thick films we discuss are of nanometric thickness. We show that the nanofluidic transport inside such films depends on several nanometric length scales, leading to a rich macroscopic response of the whole system. In particular, we demonstrate that the zeta-potential of non-thick coatings becomes a property, defined by the relative values of  their thickness, the Brinkman and Debye screening lengths, and of another electrostatic length $\ell$, which depends on the volume charge density, but not on the salt concentration.

In Sec.~\ref{sec:general} we give basic principles, brief summary of known relationships, and formulate the
problem. Solutions to electro-osmotic velocities and zeta-potentials are derived in Sec.~\ref{Surface_slip}. We illustrate the theory and validate it numerically in Sec.~\ref{sec:results}. Implications for the use of non-thick porous films to enhance electro-osmotic flows at different salt concentration are discussed in Sec.~\ref{sec:salt}, followed by concluding remarks in Sec.~\ref{sec:conclusion}

\section{Model, governing equations, and summary of known relationships }\label{sec:general}

\begin{figure}[tbp]
\begin{center}
\includegraphics[width=8cm]{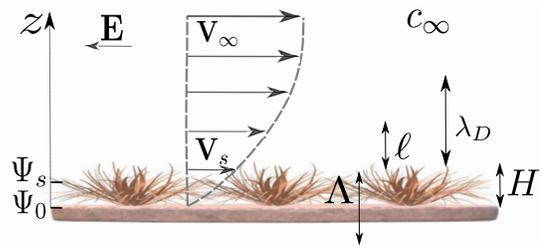}
\end{center}
\par
\vspace{-0.4cm}
\caption{Permeable non-thick coating of thickness $H$ and positive volume charge density $\varrho$ in contact with a bulk electrolyte solution of concentration $c_{\infty}$. The wall, $\Psi_0$, and surface, $\Psi_s$, electrostatic potentials are build up self-consistently and defined, besides $H$, by two lengths, $\lambda_D \propto c_{\infty}^{-1/2}$ and $\ell \propto \varrho^{-1/2}$.  The application of a tangential electric field, \boldmath{$E$}, leads to an electro-osmotic flow of solvent that depends on the Brinkman screening length $\Lambda$. The velocity at the surface of a porous film is \boldmath{$V_s$}, and that in the bulk is \boldmath{$V_{\infty}$}.} 
\label{fig:Fig1}
\end{figure}

The system geometry is shown in Fig.~\ref{fig:Fig1}. The properties of the sketched heterogeneous supported film are, of course, related to its internal structure and can be evaluated in specific situations,  but here we do not try to solve the problem at the scale of the individual pores. Instead, motivated by the theory of heterogeneous media~\cite{Markov:2000,Torquato:2002}, we replace such a real coating  by an imaginary homogeneous one, which `macroscopically' behaves in the same way and possess effective properties, such as a volume charge density or a hydrodynamic permeability. Thus, we consider a  homogeneous permeable film of a thickness $H$, which sets a length scale for our problem, of a volume charge density $\varrho$, taken positive without loss of generality.

The film is in contact with a semi-infinite 1:1 electrolyte of bulk ionic concentration $c_{\infty}$, permittivity $\varepsilon$, and dynamic
viscosity $\eta$. Ions obey Boltzmann distribution, $c_{\pm}(z)=c_{\infty}\exp (\mp \psi (z))$, where $\psi (z)=\textsl{e} \Psi (z)/(k_{B}T)$ is the dimensionless electrostatic potential, $\textsl{e}$ is the elementary positive charge, $k_{B}$ is the Boltzmann constant, $T$ is a temperature, and the upper (lower) sign corresponds to the cations (anions). In the bulk, i.e. far away from the coating ($z \to \infty$), an electrolyte solution is electro-neutral, $c_{\pm}(z)=c_{\infty}$, and $\psi (\infty) \to 0$.
The inverse Debye screening length of an electrolyte solution, $\kappa \equiv \lambda_D^{-1}$, is defined as usually, $\kappa^2=8 \pi \ell_B c_{\infty}$, with the Bjerrum length $\ell_B=\dfrac{\textsl{e}^2}{\varepsilon k_BT}$. The Debye length defines a new (electrostatic) length scale and is the measure of the thickness of the outer diffuse layer, where the local electro-neutrality is broken. We emphasize that it is independent on the film charge.

The system subjects to a weak tangential electric field $E$, so that in steady state $\psi (z)$ is independent of the fluid flow.  For our geometry the concentration gradients at every location are perpendicular to  the direction of the flow, it is therefore legitimate to  neglect advection. Consequently, the dimensionless velocity of an electro-osmotic flow, $%
v(z)=\displaystyle\frac{4\pi \ell _{B}\eta }{\textsl{e} E}V(z)$, satisfies the generalized Stokes equation~\cite{note2}
\begin{equation}\label{eq:Stokes}
v_{i,o}^{\prime \prime }-\mathcal{K}^{2}  v_{i,o} \Theta (H-z)=\psi _{i,o}^{\prime
\prime }+\kappa ^{2}\rho \Theta (H-z),
\end{equation}%
where $^{\prime}$ denotes $d/d z$, with the index $\{i, o\}$ standing for
``in'' $(z \leq H)$ and ``out'' $(z \geq H)$,  $\Theta(z)$ is the Heaviside step function, $\mathcal{K} = \Lambda^{-1}$ is the inverse Brinkman length, and $\rho = \dfrac{\varrho}{2 \textsl{e} c_{\infty}}$.  For small volume charge and/or high electrolyte concentration $\rho$ is small and below we refer such coatings to as weakly charged. For large volume charge and/or dilute electrolyte solutions $\rho$ is large and we term these films strongly charged. The Brinkman length can theoretically vary from $0$ to $\infty$. In the latter (idealized) case an additional dissipation in the porous film is neglected. If so, the hydrodynamic permeability of the porous film reaches its highest possible limit and $\propto H^2$~\cite{vinogradova.oi:2020}. In the former case of vanishing $\Lambda$ the additional dissipation inside the coating is so high that the porous film permeability ($\propto \Lambda^{2}$) tends to zero, i.e. the inner flow is fully suppressed. In reality, however $\Lambda$ is finite and defined by the parameters of the  porous film, such, for example, as volume fraction of the solid, size of the pores, and their geometry.

At the wall we apply a classical no-slip condition, $v_{0}=v_i(0)=0$, and at the surface the condition of continuity of velocity, $v_i (H) = v_o (H)$, and shear rate, $v_i^{\prime } (H) = v_o^{\prime } (H)$, is imposed. Far from the surface, the solution of Eq.\eqref{eq:Stokes} should satisfy $v_o^{\prime} \to 0$ at $z \to \infty$ to provide a plug flow. The velocity $v_{\infty}$ at $z \to \infty$ is constant and equal to $- \zeta$.

The dimensionless zeta-potential, $\zeta = \textsl{e} Z / (k_B T) $ is given by~\cite{vinogradova.oi:2020,silkina.ef:2020b}
\begin{equation}\label{eq:zeta_potential}
\zeta = \psi_s  - v_s,
\end{equation}
where $\psi_s = \textsl{e} \Psi_s /(k_B T) $ and $v_s= v(H)$. Note that $- v_s$ represents the velocity jump inside the porous film. Any situation where the value
of the tangential component of velocity appears to be different
from that of the solid surface is normally termed slip~\cite{vinogradova.oi:2011}. Therefore, $-v_s$ represents the (positive definite) slip velocity of liquid at the film surface, $z = H$. As a side note,  since the film with an outer diffuse layer is much thinner than any of the macroscopic dimensions, the bulk liquid also appears to slip, but with the velocity $-v_{\infty}$.
By this reason in colloid science  $-v_{\infty}$ is often termed an apparent electro-osmotic slip velocity.~\cite{delgado.av:2007}.  To distinguish between real and apparent slip, and recalling that  $- v_{\infty} = \zeta$, below we will refer  this (dimensionless) apparent slip to as $\zeta$.

\citet{silkina.ef:2020b} carried out calculations in the limit of zero and infinite $\mathcal{K} H$ and
 concluded that for films of an arbitrary thickness at  $\mathcal{K}H \to 0$

\begin{equation}  \label{eq:v_s_large_BL}
- v_s \simeq  \Delta \psi + \frac{\rho (\kappa H)^{2}}{2}, \, \zeta \simeq \psi_0 + \frac{\rho (\kappa H)^{2}}{2},
\end{equation}
and
\begin{equation} \label{eq:v_s_thin_small_BL}
- v_{s} \simeq   0, \, \zeta \simeq \psi_s
\end{equation}
when $\mathcal{K}H \to \infty$. Here $\psi_0 = \textsl{e} \Psi_0 /(k_B T) = \psi(0)$ is the wall potential, $\psi_s = \psi(H)$ is the surface potential, and $\Delta \psi = \psi_0 - \psi_s$ is the drop of the electrostatic potential in the coating.

Eqs.\eqref{eq:v_s_large_BL} and \eqref{eq:v_s_thin_small_BL} represent rigorous upper and lower bounds that constrain the attainable values of  slip velocity and zeta-potential. In many regimes, however, these bounds are not close enough to obviate the need for  calculations  flows over
porous surfaces of a finite $\mathcal{K}H$.

It follows from Eq.(\ref{eq:Stokes}) that to calculate electro-osmotic velocity we have to find the distribution of electrostatic potentials that satisfy the nonlinear Poisson-Boltzmann equation
\begin{equation}\label{eq:PB_io}
\psi_{i, o}^{^{\prime\prime}} = \kappa^2 \left(\sinh\psi_{i, o} -\rho\Theta
\left(H - z \right)\right) ,
\end{equation}
and to obtain simple expressions for $\psi_0$, $\psi_s$, and $\Delta \psi$.
We assume that the wall is uncharged, $\psi_i^{'}(0) = 0$, and set  $\psi_i(H) = \psi_o(H)$ and
$\psi_i^{'}(H) = \psi_o^{'}(H)$ at the surface of the coating.

The solution of Eq.(\ref{eq:PB_io}) satisfying $\psi_o \rightarrow$ 0 and $\psi_o^{'} \rightarrow$ 0 at $z \rightarrow \infty$ is the same as for an impenetrable wall of the same $\psi_s$~\cite{andelman.d:2006book}
\begin{equation}\label{eq:PB_out}
\psi_{o} (z) = 4 \artanh \left[ \gamma e^{-\kappa (z-H)}\right],
\end{equation}
where $\gamma =\tanh\dfrac{\psi_s}{4}$.

\begin{table}
  \centering
  \caption{Various limits of electrostatic ``thickness'' for a porous film of a (geometric) thickness $H$.}
  \renewcommand{\baselinestretch}{2}\normalsize
  \label{tbl:Ethickness}
  \begin{tabular}{l|c|c}
  \multicolumn{3}{c}{} \\
  \hline
     Electrostatic & Weakly charged & Highly charged\\
  thickness    & films ($\rho \ll 1$)  & films ($\rho \gg 1$)  \\
  \hline
   thick & $\kappa H \gg 1$  &  $\kappa H\sqrt{\rho}  \gg 1$  \\
    \hline
   non-thick &  $\kappa H \ngg 1$  & $\kappa H\sqrt{\rho}  \ngg 1$ \\
  \hline
  thin & $\kappa H \ll 1$  &  $\kappa H\sqrt{\rho}  \ll 1$  \\
    \hline
        \end{tabular}
\end{table}

Ions of an outer electrolyte can permeate inside the porous film, giving rise to their homogeneous equilibrium distribution in the system, with the enrichment of anions in the film. When $ \kappa H (1+\rho ^{2})^{1/4} \gg 1$ the film becomes thick compared to the inner diffuse layer, with an extended `bulk' electro-neutral region (where intrinsic coating charge is completely screened by absorbed electrolyte ions, is formed). The potential in this region is usually referred to as the Donnan potential, $\psi_D$. Note that Eq.\eqref{eq:PB_io} immediately suggests that $\psi_D =\arsinh(\rho )$ since in the electro-neutral area $\psi_{i}^{^{\prime\prime}}$ vanishes.
A systematic treatment of the influence of the Brinkman length on the zeta-potential of thick films was contained in a paper published by~\citet{vinogradova.oi:2020}. Here we will focus on the case of films of
\begin{equation}\label{eq:criterion}
   \kappa H (1+\rho ^{2})^{1/4} \ngg 1
\end{equation}
that can be termed non-thick. For weakly charged films of $\rho \ll 1$ this implies that $\kappa H = O(1)$ or smaller. The more interesting strongly charged coatings of $\rho \gg 1$ can be considered as non-thick when $\kappa H\sqrt{\rho}  \ngg 1$ and are thin when $\kappa H\sqrt{\rho}  \ll 1$. For convenience in Table~\ref{tbl:Ethickness} we give a summary of criteria defining different limits for an electrostatic thickness.

 Non-thick films do not contain an electro-neutral portion, where the intrinsic volume charge is fully screened by absorbed ions. Consequently, their $\psi_0$ given by~\cite{silkina.ef:2020b}
\begin{equation}\label{eq:psi_0}
\psi_0 \simeq \ln\left[ \frac{2+(\rho \kappa H)^{2}+\rho \kappa H \sqrt{ 4 + (\kappa H)^{2}(1 + \rho^2)}}{2+\rho (\kappa H)^{2} } \right].
\end{equation}
is smaller than the Donnan potential.

The surface potential, $\psi_s$, and the potential drop in the film,  $\Delta \psi = \psi_0 - \psi_s$, are related to $\psi_0$ as~\cite{silkina.ef:2019,silkina.ef:2020b}
\begin{equation}\label{eq:psi_s_thick}
\psi_s \equiv \psi_0 - \frac{\cosh \psi_0 - 1}{\rho}, \, \Delta \psi = \frac{\cosh \psi_0 - 1}{\rho}.
\end{equation}

The inner $\psi$-profile of a non-thick film is given by~\cite{silkina.ef:2020b}
\begin{equation}\label{eq:PB_in1}
\psi_{i}(z)\simeq \psi_0 - \dfrac{\rho}{2} \left(\kappa z\right)^2 \left[ 1 - \mathcal{F}\right],
\end{equation}
where

\begin{equation}\label{eq:F}
  \mathcal{F}= \dfrac{\sinh \psi_0}{\rho}
\end{equation}
represents the fraction of the screened film intrinsic charge at $z = 0$.

The surface potential is then
\begin{equation}\label{eq:psi_s_thin}
\psi_s \simeq \psi_0 - \dfrac{\rho}{2} \left(\kappa H\right)^2 \left[ 1 - \mathcal{F} \right].
\end{equation}

\section{Electrostatic potentials vs. zeta-potential}\label{Surface_slip}

The expression for an outer velocity can be written as~\cite{silkina.ef:2020b,vinogradova.oi:2020}
\begin{equation}\label{eq:v_inf}
 v_{o}(z) = v_s  + \psi_{o}(z) - \psi_{s},
\end{equation}
where $\psi_o$ is given by Eq.\eqref{eq:PB_out} and $\psi_s$ obeys Eq.\eqref{eq:psi_s_thin}. Therefore, in order to obtain a detailed information concerning zeta-potential a calculation of $v_s$ arising due to the inner flow is required.

 We have calculated the inner velocity profile by solving Eq.(\ref{eq:Stokes}) with $\psi_i$ satisfying Eq.(\ref{eq:PB_in1}) and prescribed boundary conditions, and obtained that $v_i$ is given by

\begin{widetext}
\begin{equation}\label{eq:EO_in_thin}
v_{i}= \left(\rho \left( \frac{\kappa }{\mathcal{K}}\right) ^{2} - \frac{2 \Delta \psi}{\left( \mathcal{K} H \right)^{2}}\right) \left( e^{- \mathcal{K}z} - 1 \right) +  \frac{\sinh\mathcal{K}z}{\cosh\mathcal{K}H} \left[ \left( \rho \left( \frac{\kappa }{\mathcal{K}}\right) ^{2} - \frac{2 \Delta \psi}{\left( \mathcal{K} H \right)^{2}}  \right) e^{-\mathcal{K}H} - \frac{2 \Delta \psi}{\mathcal{K} H} \right].
\end{equation}

so that at the surface

\begin{equation} \label{eq:v_s_thin}
v_s= \left[ \rho \left(\kappa H\right) ^{2} - 2 \Delta \psi \right] \frac{ (1 + \tanh\mathcal{K}H ) e^{-\mathcal{K}H} -1 }{(\mathcal{K}H)^2} - 2 \Delta \psi \frac{\tanh\mathcal{K}H}{\mathcal{K} H},
\end{equation}%
\end{widetext}
Eq.\eqref{eq:v_s_thin} can be used for any values of $\rho$ and $\mathcal{K}H$, and in the limits of $\mathcal{K}H \to 0$ and $\infty$ reduces to Eqs.\eqref{eq:v_s_large_BL} and \eqref{eq:v_s_thin_small_BL}.

When $\mathcal{K}H $ is small, Eq.\eqref{eq:v_s_thin} can be expanded about $\mathcal{K}H = 0$, and to second order we obtain
\begin{equation}  \label{eq:v_s_large_BL_next}
- v_s \simeq \Delta \psi \left( 1 - \frac{(\mathcal{K}H)^2}{4}\right) + \frac{\rho (\kappa H)^{2}}{2} \left( 1 - \frac{5 (\mathcal{K}H)^2}{12}\right)
\end{equation}
The first term in Eq.\eqref{eq:v_s_large_BL_next} is associated with the reduction of the
potential, $\Delta \psi$, in the porous film, but also
depends on $\mathcal{K}H$.  The second term is associated with a body
force $\rho \kappa^2$ that drives the inner flow. Both terms reduce with $\mathcal{K}H$ leading to deviations from the upper value of $-v_s$ defined by Eq.\eqref{eq:v_s_large_BL}.
Using then  \eqref{eq:zeta_potential} we conclude that the $\zeta$-potential can be approximated by
\begin{equation}\label{eq:zeta_large_BL_oiv}
\zeta \simeq \psi_0 -  \Delta \psi \frac{(\mathcal{K}H)^2}{4} + \frac{\rho (\kappa H)^{2}}{2}\left( 1 - \frac{5 (\mathcal{K}H)^2}{12}\right)
\end{equation}

Expanding $v_s$ in Eq.\eqref{eq:v_s_thin} at large $\mathcal{K}H $ we find
\begin{equation} \label{eq:v_s_thin_small_BL_next}
- v_{s} \simeq  \frac{ 2 \Delta \psi }{\mathcal{K} H} + \rho \left( \frac{\kappa }{\mathcal{K}}\right)^{2}
\end{equation}
Eq.\eqref{eq:v_s_thin_small_BL_next} indicates that $v_s$ is a superposition of a flow that is linear in $\Delta \psi$ and of a plug flow,
$\rho \left( \dfrac{\kappa }{\mathcal{K}}\right)^{2}$.
Then it follows from Eq.\eqref{eq:zeta_potential} that
\begin{equation}\label{eq:zeta_small_BL_next}
\zeta \simeq \psi_s + \frac{ 2 \Delta \psi }{\mathcal{K} H} + \rho \left( \frac{\kappa }{\mathcal{K}}\right)^{2}
\end{equation}

Thus, our treatment clarifies that at a given $\mathcal{K}H$, a slip velocity $-v_s$ (and a consequent $\zeta$) can be enhanced  by generating  larger $\Delta \psi$ and/or when $\rho (\kappa H)^{2}$ is large. When both are small, $-v_s \simeq 0$ and $\zeta \simeq \psi_s$.

The value of $\psi_0$ can be generally calculated from Eq.\eqref{eq:psi_0}, which then allows to find $\psi_s$ and $\Delta \psi$ from Eq.\eqref{eq:psi_s_thick}. Using standard manipulations we derive
\begin{widetext}
\begin{equation}\label{eq:delta_phi_oiv}
  \Delta \psi \simeq \dfrac{\rho (\kappa H)^2}{2 + \rho (\kappa H)^2}  \left( 1 + \dfrac{ (\kappa H)^2 (1 - \rho) - \kappa H\sqrt{4 + (\kappa H)^2 (1 + \rho^2)} }{ 2 + (\rho \kappa H)^2 + \rho \kappa H \sqrt{4 + (\kappa H)^2 (1 + \rho^2)}}\right)
\end{equation}
and

\begin{equation}\label{eq:Naren_F}
\mathcal{F} \simeq \frac{2 + (\rho \kappa H)^{2} }{2\rho + (\rho \kappa H)^{2}} - \frac{ 2}{\rho \left( 2+(\rho \kappa H)^{2} + \rho \kappa H \sqrt{4 + (\kappa H)^{2}(1 + \rho^2)}\right) }
\end{equation}
\end{widetext}

These two last  equations are expected to be very accurate, but are quite cumbersome. Fortunately, in some limits they can be dramatically simplified leading to very simple analytic solutions for $\zeta$. We discuss now separately two limits,
depending on how strong the dimensionless volume charge density is.

\subsection{Weakly charged coatings ($\rho \ll 1$)}\label{sec:small_rho}

At small $\rho$ one can expand $\psi_0$ given by Eq.\eqref{eq:psi_0} into a series about $\rho = 0$, and we conclude that a sensible approximation for $\psi_0$ should be

\begin{equation}\label{eq:psi_0_s_small_rho_oiv}
 \psi_0 \simeq   \rho \kappa H \dfrac{\sqrt{4+(\kappa H)^2} - \kappa H}{2}.
 \end{equation}
Note that $\psi_0$ is linear in $\rho$, but is a non-linear function of $\kappa H$ since to derive Eq.\eqref{eq:psi_0_s_small_rho_oiv} we do not make an additional assumption that $ \kappa H \ll 1$. Consequently, this and following equations of this subsection should be valid even when $\kappa H = O(1)$.

Expanding Eq.\eqref{eq:psi_s_thick} at small $\psi_0$ and substituting Eq.\eqref{eq:psi_0_s_small_rho_oiv} we obtain
\begin{equation}\label{eq:smallrho_dpsi}
\Delta \psi \simeq  \dfrac{\psi_0^2}{2 \rho} \simeq \dfrac{\rho(\kappa H)^2 (\sqrt{4+(\kappa H)^2} - \kappa H)^2}{8}  ,
\end{equation}
which together with \eqref{eq:psi_0_s_small_rho_oiv} leads to
\begin{equation}\label{eq:phissmallrho}
  \psi_s \simeq   \dfrac{\rho \kappa H}{4} \left( \sqrt{4+(\kappa H)^2}(2 + (\kappa H)^2) - \kappa H (4 + (\kappa H)^2)  \right)
\end{equation}
Note that imposing the condition of small $\kappa H$ one can easily recover the known result of the linearized Poisson-Boltzmann theory (see Appendix~\ref{Ap:linear})
\begin{equation}\label{eq:small_rho_kappa_H}
    \psi_0 \simeq  \rho \kappa H \left(1 - \frac{\kappa H}{2} \right), \, \psi_s \simeq  \rho \kappa H \left(1 - \kappa H \right),
\end{equation}
which suggests that the $\psi$-profile is almost constant throughout a weakly charged thin film.

Expanding Eq.\eqref{eq:F} at small $\psi_0$ and using Eq.\eqref{eq:psi_0_s_small_rho_oiv}  we get

\begin{equation}\label{eq:F_small_rho}
\mathcal{F} \simeq \dfrac{\psi_0}{\rho} = \dfrac{\kappa H (\sqrt{ 4 + (\kappa H)^2} - \kappa H)}{2} + O(\rho^2)
\end{equation}
We remark that in this low $\rho$ regime to leading order $\mathcal{F}$ does not depend on $\rho$, and is finite even if $\rho \to 0$, where $\psi_0 \simeq 0$.  At first sight this is somewhat surprising, but we recall that our dimensionless charge density is introduced by dividing the real one by the salt concentration, so that a nearly vanishing $\rho$ simply implies that the (non-thick) film is enriched by counter-ions that partly screen its intrinsic charge.

It is clear that $\psi_0$, $\psi_s$, and $\Delta \psi$ are small, so is $v_s$ given by Eq.\eqref{eq:v_s_large_BL_next}. Consequently, $\zeta$ is also generally small and we do not discuss it here in detail. However, it would be worthwhile to mention that an upper bound on $\zeta$ in this case is

\begin{equation}
  \zeta \simeq \rho \kappa H \dfrac{\sqrt{4+(\kappa H)^2} }{2},
\end{equation}
which together with \eqref{eq:phissmallrho} gives $\zeta/\psi_s \simeq 2 + \kappa H \left(1 - \dfrac{(\kappa H)^2}{8}\right)$. Thus, the electro-osmotic flow in the bulk can potentially be enhanced in more than two times compared to the Smoluchowski case.

\subsection{Strongly charged coatings ($\rho \gg 1$)}\label{sec:theory_large_rho}

For strongly charged coatings Eq.\eqref{eq:psi_0} reduces to~\cite{silkina.ef:2020b}

\begin{equation}\label{eq:psi_0_thin_modified3}
\psi_0 \simeq 2 \arsinh\left(\dfrac{\rho \kappa H}{2}\right)  - \ln \left(1 +\dfrac{\rho (\kappa H)^2}{2}\right).
\end{equation}

Straightforward calculations show that Eqs.\eqref{eq:delta_phi_oiv} and \eqref{eq:Naren_F} can be transformed to

\begin{equation}\label{eq:delta_phi_oiv2}
  \Delta \psi \simeq \dfrac{\rho (\kappa H)^2}{2 + \rho (\kappa H)^2},
\end{equation}

\begin{equation}\label{eq:Naren_F_oiv}
\mathcal{F} \simeq \dfrac{2 + (\rho \kappa H)^{2} }{2\rho + (\rho \kappa H)^{2}} \simeq  \Delta \psi + \dfrac{2  }{\rho (2 + \rho( \kappa H)^{2})},
\end{equation}
indicating that $\Delta \psi \simeq \mathcal{F}$ when $\rho$ is large.

Two limits can now be distinguished depending on the
value of $\rho (\kappa H)^{2}$.

\subsubsection{The limit of $\rho (\kappa H)^{2} \ll 1$}\label{sec:small_rho_kH2}

We recall that since $\rho$ is large, the film becomes thin when $ \kappa H \sqrt{\rho}\ll 1$ (see Table~\ref{tbl:Ethickness}). Therefore,
such a situation is possible only for films that are truly or relatively thin (below we refer the latter to as \emph{quasi-thin}).
We further remark that in this limit
the first term in Eq.\eqref{eq:psi_0_thin_modified3} dominates, so that it can be further simplified to give
\begin{equation}\label{eq:psi_0_thin_modified3_oiv}
\psi_0 \simeq 2 \arsinh\left(\dfrac{\rho \kappa H}{2}\right)  - \dfrac{\rho (\kappa H)^2}{2}
\end{equation}
In turn, Eq.\eqref{eq:delta_phi_oiv2} reduces to

\begin{equation}\label{eq:delta_phi_smallrhokH}
  \Delta \psi \simeq  \dfrac{\rho (\kappa H)^2}{2},
\end{equation}
leading to
\begin{equation}\label{eq:psi_s_thin_modified3_oiv}
\psi_s \simeq 2 \arsinh\left(\dfrac{\rho \kappa H}{2}\right)  - \rho (\kappa H)^2
\end{equation}

When $\rho \kappa H$ is small, Eqs.\eqref{eq:psi_0_thin_modified3_oiv} and \eqref{eq:psi_s_thin_modified3_oiv} reduce to Eqs.\eqref{eq:small_rho_kappa_H}, which implies  that they can also be employed when $\rho$ is small, provided $\kappa H$ is not large.

From Eq.\eqref{eq:zeta_large_BL_oiv} we then find that for small $\mathcal{K} H$ the zeta-potential can be approximated as
\begin{equation}\label{eq:zeta_large_BL_oiv2}
\zeta \simeq 2 \arsinh\left(\dfrac{\rho \kappa H}{2}\right) - \frac{\rho (\kappa H)^{2}(\mathcal{K}H)^2}{3},
\end{equation}
which leads to
\begin{equation}\label{eq:zeta_smallK_oiv}
\zeta \simeq 2 \arsinh\left(\dfrac{\rho \kappa H}{2}\right)
\end{equation}
when $\mathcal{K}H \to 0$. However, using \eqref{eq:psi_s_thin_modified3_oiv} we obtain $\zeta - \psi_s \leq \rho (\kappa H)^2,$ which is small in this limit. Therefore, our asymptotic arguments suggest that even in the case of extremely large hydrodynamic permeability of the porous layer, the difference between $\zeta$ and $\psi_s$ cannot be  significant. Thus a knowledge of $\psi_s$ should be sufficient to provide a realistic evaluation of $\zeta$ (and vice versa). Nevertheless, for completeness we mention that at large $\mathcal{K} H$ from \eqref{eq:zeta_small_BL_next} one can obtain

\begin{equation}\label{eq:zeta_largeK_oiv}
\zeta \simeq 2 \arsinh\left(\dfrac{\rho \kappa H}{2}\right) +  \rho \left( \frac{\kappa }{\mathcal{K}}\right)^{2} (1 + \mathcal{K} H - (\mathcal{K} H)^2),
\end{equation}
which tends to $\psi_s$ given by Eq.\eqref{eq:psi_s_thin_modified3_oiv} when $\mathcal{K} H \to \infty$.

\subsubsection{The limit of $\rho (\kappa H)^{2} \gg 1$}

This limit is close to, but weaker of, the condition for a thick film $\kappa H \sqrt{\rho} \gg 1$ (see Table~\ref{tbl:Ethickness}). So, we can term these films \emph{quasi-thick}.

For large $\rho (\kappa H)^2$, Eqs.\eqref{eq:psi_0_thin_modified3} and \eqref{eq:delta_phi_oiv2} can be further simplified to
\begin{equation}\label{eq:psi_0_large_rho_large_rho_hh}
\psi_0 \simeq \ln(2 \rho) - \dfrac{2}{\rho (\kappa H)^2}, \,  \Delta \psi \simeq  1 - \dfrac{2}{\rho (\kappa H)^2}
\end{equation}
which gives the same $\psi_s$ as for thick films~\cite{silkina.ef:2020,vinogradova.oi:2020}
\begin{equation}\label{eq:psi_s_large_rho_large_rho_hh}
\psi_s \simeq \ln(2 \rho) - 1
\end{equation}

Substituting \eqref{eq:psi_0_large_rho_large_rho_hh} into \eqref{eq:zeta_large_BL_oiv} in the limit $\mathcal{K}H \ll 1$ we obtain

\begin{equation}\label{eq:zeta_large_BL_large_rho_large_rho_hh_modified}
\zeta \simeq  \ln(2 \rho) + \frac{\rho (\kappa H)^{2}}{2} \left( 1 - \dfrac{5 (\mathcal{K} H)^{2}}{12} \right),
\end{equation}
which suggests that for quasi-thick films $\zeta$ can become very large and significantly exceeds $\psi_s$.

Substitution of Eqs.\eqref{eq:psi_0_large_rho_large_rho_hh} and \eqref{eq:psi_s_large_rho_large_rho_hh} into \eqref{eq:zeta_small_BL_next} for $\mathcal{K}H \gg 1$ leads to

\begin{equation}\label{eq:zeta_small_BL_large_rho_large_rho_hh_modified}
\zeta \simeq   \ln\left(2 \rho \right) - 1 + \rho \left( \frac{\kappa }{\mathcal{K}} \right)^{2}   + \dfrac{2}{\mathcal{K} H}
\end{equation}

\section{Numerical results and discussion}\label{sec:results}

It is of considerable interest to compare exact numerical data with our analytical theory and to determine the regimes of validity of asymptotic results.
Here we first present results of  numerical solutions of Eq.\eqref{eq:PB_io} with prescribed boundary conditions, using the collocation method~\cite{bader.g:1987}. We then solve numerically the system of Eqs.\eqref{eq:Stokes} and \eqref{eq:PB_io}. The exact numerical solutions will be presented together with calculations from the asymptotic approximations derived in Sec.\ref{Surface_slip}.

\begin{figure}[h]
\begin{center}
\includegraphics[width=1\columnwidth]{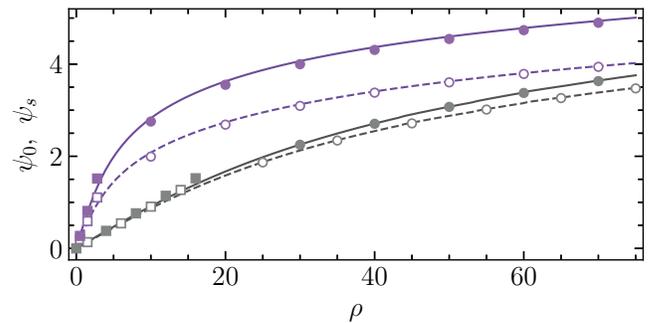}
\end{center}
\caption{Potentials at wall (solid lines) and surface (dashed)  as a function of $\rho$ computed for fixed $\kappa H = 0.8$ (upper set of curves) and $\kappa H = 0.1$ (lower curves). Filled and open squares illustrate calculations from Eqs.\eqref{eq:psi_0_s_small_rho_oiv} and \eqref{eq:phissmallrho}. Filled and open circles are obtained using Eqs.\eqref{eq:psi_0_large_rho_large_rho_hh} and \eqref{eq:psi_s_large_rho_large_rho_hh}. Filled and open triangles correspond to Eqs.\eqref{eq:psi_0_thin_modified3_oiv} and \eqref{eq:psi_s_thin_modified3_oiv}. Dash-dotted lines show $\psi_0$  from Eq.\eqref{eq:psi_0_LP}. } \label{fig:Fig2}
\end{figure}

\begin{figure}[h]
\begin{center}
\includegraphics[width=1\columnwidth]{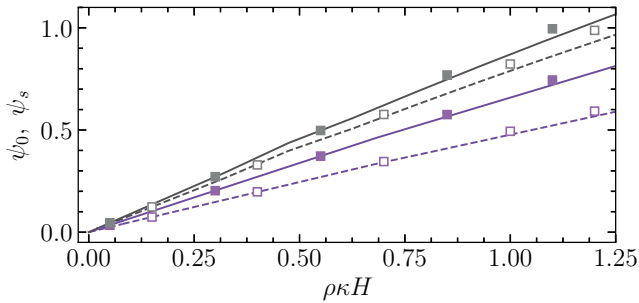}
\end{center}
\caption{The data sets for $\psi_0$ and $\psi_s$ obtained at smaller values of $\rho$ reproduced from Fig.~\ref{fig:Fig2} and plotted as a function of $\rho \kappa H$. The upper set of curves and symbols shows $\kappa H = 0.1$, the lower one corresponds to $\kappa H = 0.8$.
 }\label{fig:Fig3}
\end{figure}

In Fig.~\ref{fig:Fig2} we plot $\psi_0$ and $\psi_s$,  computed using $\kappa H = 0.8$ and $0.1$, as a function of $\rho$. It is well seen that for a thinner film $\psi_0 \simeq \psi_s$ up to $\rho \kappa H \simeq 3$. On increasing $\rho$ further $\Delta \psi$ increases slowly. For a thicker film of $\kappa H = 0.8$ the potential drop in the film is always finite and  $\Delta \psi$ grows much faster as $\rho$ is increased. The theoretical curves calculated from Eqs.\eqref{eq:psi_0_s_small_rho_oiv} and \eqref{eq:phissmallrho} are also included in Fig.~\ref{fig:Fig2}. The fits are quite good for $\rho \leq 2$, but at larger $\rho$ there is some discrepancy, especially for $\kappa H = 0.8$, and the theoretical potentials predicted by low $\rho$ approximations become
higher than computed. Note, however, that for $\kappa H = 0.1$ the discrepancy between a linear fit and numerical calculations is negligibly small when $\rho \leq 2$.
To examine its significance more closely, the initial portions of the $\psi$-profiles
from Fig.~\ref{fig:Fig2} are reproduced in Fig.~\ref{fig:Fig3}, but now plotted as a function of $\rho \kappa H$. An overall conclusion from this plot is that the approximations derived in Sec.\ref{sec:small_rho} are very accurate when $\rho \kappa H \leq 1$. We now return to Fig.~\ref{fig:Fig2} and focus on the large $\rho$ portions of the curves. As reported by \citet{silkina.ef:2020b}, Eq.\eqref{eq:psi_0} fits very accurately the numerical data for $\psi_0$ at any $\rho$, so does more elegant  \eqref{eq:psi_0_thin_modified3}, except for $\rho \leq 1$, where some very small discrepancy is observed. Calculations with our parameters  fully confirm this conclusion, so that we do not show these data. Instead, we include $\psi_0$ calculated from Eqs.\eqref{eq:psi_0_thin_modified3_oiv} and \eqref{eq:psi_0_large_rho_large_rho_hh} that correspond to small and large $\rho (\kappa H)^2$. It is well seen that for $\rho \geq 10$ the agreement with numerical data is excellent in both cases. Also included is $\psi_s$ from \eqref{eq:psi_s_large_rho_large_rho_hh} and \eqref{eq:psi_s_thin_modified3_oiv}, and we see that these asymptotic approximations coincide with the numerical data.

\begin{figure}[h]
\begin{center}
\includegraphics[width=1\columnwidth]{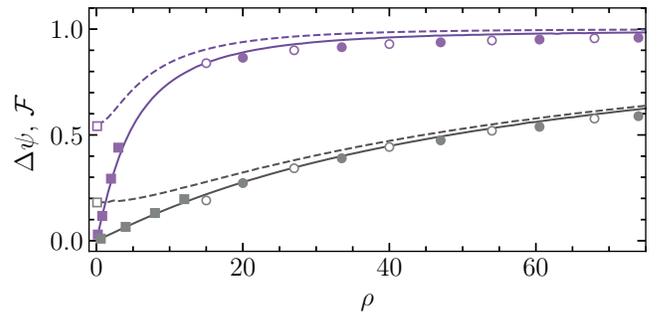}
\end{center}
\caption{ $\Delta\psi$ (solid curves) and $\mathcal{F}$ (dashed curves)  vs $\rho$ computed using $\kappa H = 0.8$ (top) and $0.2$ (bottom). Filled and open squares are calculations from Eqs.~\eqref{eq:smallrho_dpsi} and \eqref{eq:F_small_rho}.
Filled and open circles correspond to Eqs.~\eqref{eq:delta_phi_oiv2} and \eqref{eq:Naren_F_oiv}. }\label{fig:Fig4}
\end{figure}

Fig.~\ref{fig:Fig4} shows the electrostatic potential drop, $\Delta \psi$, inside the film computed for $\kappa H = 0.2$ and 0.8 as a function of $\rho$.
The degree of screened intrinsic  charge at the wall, $\mathcal{F}$, calculated numerically for the same values of $\kappa H$ is also plotted.
It is seen that $\Delta \psi$ first increases linearly with $\rho$ and, when  $\rho$ is getting sufficiently large, slowly approaches to unity for a film of $\kappa H = 0.8$. However, in a chosen interval of $\rho$ the potential drop of a thinner film of $\kappa H = 0.2$ still continues to grow, although weakly and nonlinearly. It can be seen that the linear portions of the curves are well fitted by Eq.~\eqref{eq:smallrho_dpsi}, and the nonlinear ones are reasonably well described by Eq.\eqref{eq:delta_phi_oiv2}. Also included in Fig.~\ref{fig:Fig4} are the curves for $\mathcal{F}$ computed using the same values of $\kappa H$. For strongly charged coatings $\mathcal{F} \simeq \Delta \psi$, confirming predictions of Eq.\eqref{eq:Naren_F_oiv}. When $\rho = 0$, $\mathcal{F}$ is finite and its value is given by \eqref{eq:F_small_rho}. This equation also predicts a parabolic growth of $\mathcal{F}$ at small $\rho$, which is well seen in Fig.~\ref{fig:Fig4}.

\begin{figure}[h]
\begin{center}
\includegraphics[width=1\columnwidth]{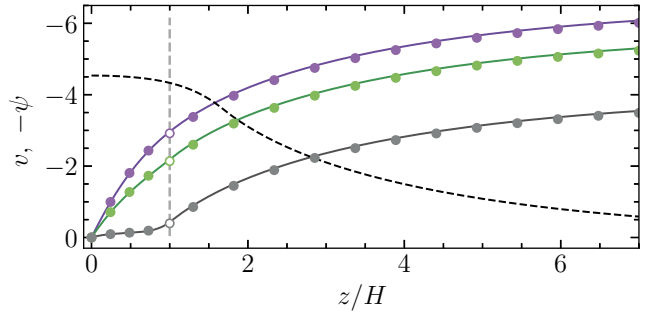}
\end{center}
\caption{The profiles $v$ computed using $\kappa H = 0.3$ and $\rho = 50$ with $\mathcal{K}H=0.1, 1,$ and 5 (solid curves from top to bottom). Dashed line shows the electrostatic potential profile taken with the negative sign, $-\psi$. Filled circles show predictions of Eqs.\eqref{eq:v_inf} and \eqref{eq:EO_in_thin}. Open circles correspond to $v_s$ calculated from Eq.~\eqref{eq:v_s_thin}.  } \label{fig:Fig5}
\end{figure}

We now turn to the electro-osmotic velocity. The velocity profiles computed using three $\mathcal{K} H$ in the range from 0.1 (small) to 5 (relatively large) are shown in Fig.~\ref{fig:Fig5}. They have been obtained using $\kappa H = 0.3$ and $\rho = 50$. Note that with these parameters $\kappa H \sqrt{\rho}  \simeq 2.1$, $\rho \kappa H = 15$, and $\rho (\kappa H)^2 = 4.5$, so in our terms we deal with a non-thick highly charged film of moderate value of $\rho (\kappa H)^2$. Also included is the computed $\psi$-profile for this film. As described in Sec.~\ref{sec:general}, the electrostatic potential of a non-thick film is generally nonuniform throughout the system. Its maximum value (at the wall) reaches about 4.5, indicating that  nonlinear electrostatic effects become significant. The theoretical curves calculated from Eq.\eqref{eq:v_inf} for $v_o$ using $v_s$ defined by Eq.\eqref{eq:v_s_thin} and from Eq.\eqref{eq:EO_in_thin} for $v_i$ coincide with the numerical data. It can be seen that on reducing $\mathcal{K} H$ the value of $-v$ increases. All outer velocity profiles are of the same shape that is set by $\psi_o$, indicating that the dramatic increase in $-v_{o}$ upon decreasing $\mathcal{K} H$ is induced by changes in $v_s$ only. At very large $z/H$ the curves for $v_o$ saturate to $v_{\infty} = - \zeta$ (not shown).

\begin{figure}[h]
\begin{center}
\includegraphics[width=1\columnwidth]{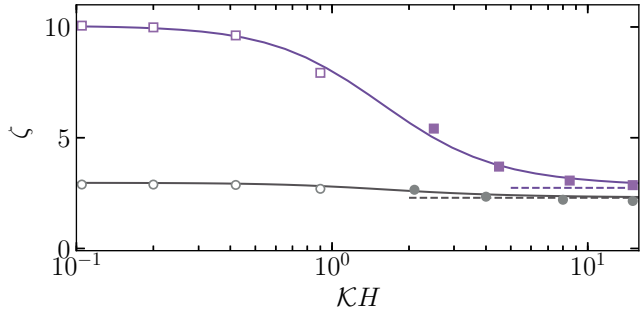}
\end{center}
\caption{Zeta potential as a function of $\mathcal{K}H$ computed for $\kappa H=0.8$ and 0.2  (solid curves from top to bottom) and $\rho = 20$. Dashed lines show $\psi_s$. Open and filled circles show predictions of Eqs.\eqref{eq:zeta_smallK_oiv} and \eqref{eq:zeta_largeK_oiv}. Open and filled squares are obtained using Eqs.\eqref{eq:zeta_large_BL_large_rho_large_rho_hh_modified} and \eqref{eq:zeta_small_BL_large_rho_large_rho_hh_modified}. } \label{fig:Fig6}
\end{figure}

Fig.~\ref{fig:Fig6} intends to indicate the range of $\zeta$ that is encountered at different $\mathcal{K} H$. For this numerical example we use films of $\kappa H = 0.2$ and 0.8, the same as in Fig.~\ref{fig:Fig4}, and explore the case of $\rho = 20$. With these parameters $\rho(\kappa H)^2 = 0.8$ and 12.8. These values differ significantly and correspond to different limits (or quasi-thin and quasi-thick films) described in Sec.~\ref{sec:theory_large_rho}, but the surface potentials, which are also shown in Fig.~\ref{fig:Fig6}, are quite close (and not small). In the chosen range of  values of $\mathcal{K}H$, which are neither too small nor quite large, zeta potentials of both films reduce strictly monotonically. We see that the value of $\zeta$ is much larger for the quasi-thick film of $\kappa H = 0.8$, where $\zeta$ can exceed $\psi_s$ in several times. For a quasi-thin film of $\kappa H = 0.2$ the zeta potential is  higher than $\psi_s$, but not much.
The parts of the $\zeta$-curves corresponding to $\mathcal{K} H \leq 1$ are well described by Eqs.\eqref{eq:zeta_smallK_oiv} and \eqref{eq:zeta_large_BL_large_rho_large_rho_hh_modified}, pointing out that this asymptotic approximations have validity well beyond the range of the original assumptions (see Sec.~\ref{Surface_slip}). When $\mathcal{K} H \geq 2$, the decay of $\zeta$ is well consistent with predictions of Eqs.\eqref{eq:zeta_largeK_oiv} and \eqref{eq:zeta_small_BL_large_rho_large_rho_hh_modified}, indicating that the latter are also valid well outside the range of their formal applicability.
As is usual, $\zeta \to \psi_s$ as $\mathcal{K} H \to \infty$.

\begin{figure}[h]
\begin{center}
\includegraphics[width=1\columnwidth]{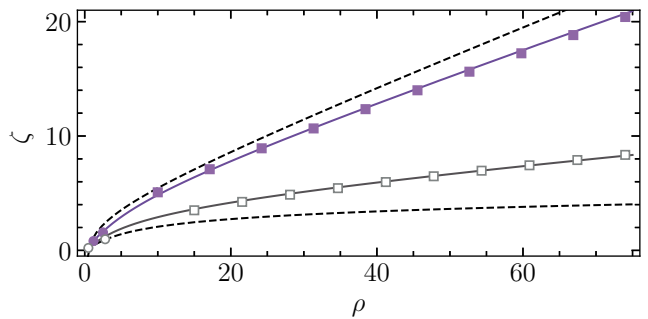}
\end{center}
\caption{Zeta-potential $\zeta$ computed for a film of $\kappa H = 0.7$  as a function of $\rho$ for fixed $\mathcal{K}H = 0.6$ and 3 (upper and lower solid curves). Filled and open circles show calculations from Eqs.\eqref{eq:zeta_large_BL_oiv2} and \eqref{eq:zeta_largeK_oiv}. Filled and open squares are obtained using Eqs.\eqref{eq:zeta_large_BL_large_rho_large_rho_hh_modified} and \eqref{eq:zeta_small_BL_large_rho_large_rho_hh_modified}. Dashed curves indicate upper and lower bounds on $\zeta$.} \label{fig:Fig7}
\end{figure}

We now fix $\kappa H = 0.5$ and compute $\zeta$ as a function of $\rho$ using $\mathcal{K} H = 0.6$ and 3. The results are plotted in Fig.~\ref{fig:Fig7}, and compared with upper and lower bounds on $\zeta$ given by Eqs.\eqref{eq:v_s_large_BL} and \eqref{eq:v_s_thin_small_BL}. The computed at finite $\mathcal{K} H$ zeta potentials are naturally confined between these two values. For small $\rho (\kappa H)^2$ we observe a rapid increase of $\zeta$ with $\rho$ that is well described by Eqs. \eqref{eq:zeta_large_BL_oiv2} and \eqref{eq:zeta_largeK_oiv}. As $\rho$ is increased, $\rho (\kappa H)^2$ is shifted to a large value and
formulas \eqref{eq:zeta_large_BL_large_rho_large_rho_hh_modified} and \eqref{eq:zeta_small_BL_large_rho_large_rho_hh_modified} become very accurate.

\section{Towards   switching  surface  and zeta potentials by salt}\label{sec:salt}

So far we have considered $\psi_s$, $\psi_0$, and $\zeta$ using dimensionless variables, such as $\rho$, $\kappa H$, $\mathcal{K}H$, and their combinations. Additional insight into the problem can be gleaned by expressing $\zeta$ as a function of characteristic length scales. These are the geometric length $H$,  the hydrodynamic one $\Lambda$, and, of course, the electrostatic length $\lambda_D$. We recall that a useful formula for 1:1 electrolyte is~\cite{israelachvili.jn:2011}
\begin{equation}\label{eq:DLength}
  \lambda_D [\rm{nm}] = \frac{0.305 [\rm{nm}]}{\sqrt{c_{\infty}}[\rm{mol/L}]},
\end{equation}
and the dependence of $\psi_s$ and $\zeta$ on $\lambda_D$ in the equations below reflects their dependence on
$c_{\infty}$. The later is often probed in electrokinetic experiments, where a decrease of both potentials with salt is observed~\cite{sobolev.vd:2017,lorenzetti.m:2016,irigoyen.j:2013}. We stress, however, that the measurements have been often conducted by using only a very narrow range of relatively large $c_{\infty}$ since existing linear theories could not provide a reasonable interpretation of data at low concentrations, where potentials are high.

It is also convenient to introduce a new electrostatic length of the problem
\begin{equation}\label{eq:ell}
 \ell = \sqrt{\dfrac{\textsl{e}}{4 \pi \ell_B \varrho}} \propto \varrho^{-1/2},
\end{equation}
which is inversely proportional to the square root of the volume charge density, but does not depend on the bulk salt concentration.

The definition of dimensionless $\rho$ can then be reformulated as
\begin{equation}\label{eq:rho}
  \rho =  \left( \dfrac{\lambda_D}{\ell}\right)^2
\end{equation}
This suggests that it is the ratio of two electrostatic length scales of the problem that determines whether coatings are weakly or strongly charged.  It is clear that an interesting ``cross-over''  behavior
must occur for some intermediate values $c_{\infty}$ that corresponds to $\lambda_D \simeq \ell$. We return to this important point below.

Condition \eqref{eq:criterion} of a non-thick film then becomes
\begin{equation}\label{eq:criterion_lengs}
\dfrac{H}{\lambda_D} \left( 1 + \left(  \dfrac{\lambda_D}{\ell}\right)^4 \right)^{1/4} \ngg 1,
\end{equation}
i.e. $H/\lambda_D \ngg 1$ for weakly charged films, and $H/\ell \ngg 1$ for highly charged films. It is instructive to mention that an electrostatic thickness of highly charged films is equal to $H/\ell$ and does not depends on salt. Therefore, such films are thin when $H/\ell \ll 1$, but weakly charged films are thin when $H/\lambda_D \ll 1$.

The combinations of dimensionless parameters that control $\psi_s$, $\psi_0$, and $\zeta$ can then be related to $\ell$ as
\begin{equation}\label{eq:rhokH}
 \rho \kappa H = \dfrac{\lambda_D H}{\ell^2}, \, \rho (\kappa H)^2 = \left( \frac{H}{\ell}\right)^2, \, \rho \left( \dfrac{\kappa}{\mathcal{K}} \right)^2 = \left( \dfrac{\Lambda}{\ell} \right)^2
\end{equation}
Eqs.\eqref{eq:rhokH} illustrate that there exist several length scales, lying always in the nanometric
range, which determine different regimes of the electro-osmotic flow. Another important conclusion from Eqs.\eqref{eq:rhokH} is that $\rho (\kappa H)^2$ and $\rho \left( \kappa/\mathcal{K} \right)^2$ do not depend on the salt concentration in the bulk. Accordingly, the dependence on salt is hidden only in $\rho \kappa H$, which is the function of $\lambda_D$.

We now present some results illustrating the role of length scales and showing that for films of a given $H/\ell$ the electrostatic regimes (of thin and thick films, or highly and weakly charged coatings) can be  tuned by the concentration of salt.
Let us now keep fixed $H = 15$ nm and consider two films, of $\ell = 5$ and 30 nm, which gives $H/\ell = 3$ and 0.5. These values of $\ell$ corresponds to $\varrho = 360$ and 10 kC/m$^3$, and we note that our larger value of $\varrho$ is close to the maximal one reported in experiments~\cite{yezek.lp:2005,braeken.l:2006}.
In our concentration range  $\lambda_D/\ell$ reduces from 60 down to 0.6 for the film of $\ell = 5$ nm, and from 10 down to 0.1 for that of
$\ell = 30$ nm. It is easy to check that with the chosen parameters both model films fall to a category of non-thick.

\begin{figure}[h]
\begin{center}
\includegraphics[width=1\columnwidth]{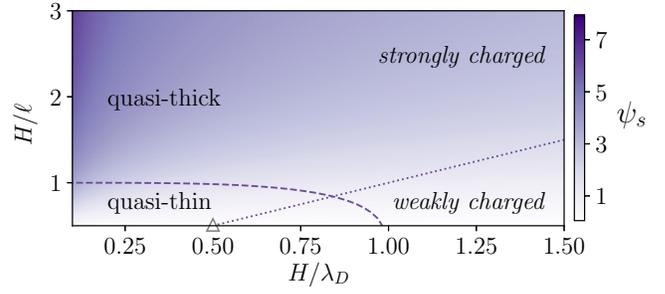}
\end{center}
\caption{Schematic representation of various electrostatic regimes for a non-thick porous film. The colorbar values of $\psi_s$ ascend from top to bottom.  The diagram is plotted in the ($H/\ell$, $H/\lambda_D$) plane. Dotted line separates the regions, where coatings obey a linear theory, and where they can only be  described using a non-linear theory. Dashed curve separates regions of quasi-thick and quasi-thin films as discussed in the text. Open triangle marks the point of $c_{\infty} = c_{\infty}^{\vartriangle}$.}\label{fig:Fig8}
\end{figure}

We begin with the treatment of $\psi_s$ obtained from the numerical solution of Eq.\eqref{eq:PB_io}. Fig.~\ref{fig:Fig8} summarize different regimes in the ($H/\ell, H/\lambda_D$) plain, where the magnitude of computed $\psi_s$ is reflected by color. The smallest and largest values of $H/\ell$ in this diagram coincide with those of the two model films specified above, and the range of $H/\lambda_D$ corresponds to $c_{\infty}$ from $10^{-6}$ to $10^{-3}$  mol/L.
It is now useful to divide the ($H/\ell, H/\lambda_D$) plane into two regions, of weakly and strongly charged films, where the above scaling expressions for $\psi_s$ approximately hold. We first remark that the conditions of weakly and highly charged coatings summarized in Table~\ref{tbl:Ethickness} coincide when $\rho = 1$. Consequently, we include in Fig.~\ref{fig:Fig8} (dotted) straight line that corresponds to $\lambda_D/\ell = 1$ (which is equivalent to $\rho = 1$) separating weakly and highly charged surfaces. When $H/\ell$ is below this line a simple linear theory can be employed.  However, for larger $H/\ell$ the Poisson-Boltzmann equation~\eqref{eq:PB_io} cannot be linearized. Apart from
this line, another crossover locus, $\dfrac{H}{\lambda_D} \left( 1 + \left(  \dfrac{\lambda_D}{\ell}\right)^4 \right)^{1/4} = 1$ (separating the quasi-thin and quasi-thick film regions~\cite{note3}) is shown by dashed curve.
 Of course, in reality at those two curves, the limiting solutions for $\psi_s$ should crossover
smoothly from one electrostatic regime to another. We can now conclude that in very dilute solutions both films are  highly charged. However, at low salt the film of  $\ell = 5$ nm is quasi-thick, but that of $\ell = 30$ nm is quasi-thin. If we increase $H/\lambda_D$ (increase $c_{\infty}$) for a film of $\ell = 30$, we move to a situation of weakly charged quasi-thin films. The intersection of the horizontal line $H/\ell = 0.5$ with the curve $\lambda_D/\ell = 1$ is marked with an open triangle and determines $c_{\infty}^{\vartriangle}$. On increasing $H/\lambda_D$ further this film becomes weakly charged quasi-thick. The film of $\ell = 5$ nm is quasi-thick for all $H/\lambda_D$ and becomes weakly charged at $c_{\infty}^{\blacktriangle}$ that is defined by the the intersection of the line $H/\ell = 3$ with $\lambda_D/\ell = 1$.

The surface potential of a highly charged quasi-thick film can be obtained using Eq.~\eqref{eq:psi_s_large_rho_large_rho_hh}
\begin{equation}\label{eq:psi_s_large_rho_hh_dimensional}
\psi_s \simeq  2 \ln \left( \dfrac{\lambda_D}{\ell} \right) + \ln 2 - 1
\end{equation}
and depends only on $\lambda_D/\ell$. However, when the highly-charged film is quasi-thin, $\psi_s$ obeys Eq.\eqref{eq:psi_s_thin_modified3_oiv} that can be rewritten as
\begin{equation}\label{eq:psi_s_length}
\psi_s \simeq 2 \arsinh\left(\dfrac{\lambda_D H}{2 \ell^2 }\right)  - \left( \dfrac{H}{\ell}\right)^2
\end{equation}
Thus, in this situation $\psi_s$ is defined by both $\lambda_D/\ell$ and $H/\ell$.

At high salt, where both films become weakly charged and quasi-thick,
to calculate $\psi_s$ one can use \eqref{eq:ohshima}, which gives
\begin{equation}\label{eq:psi_s_ohshima}
\psi_s \simeq  \dfrac{1}{2}\left( \dfrac{\lambda_D}{\ell}\right)^2 \propto \varrho c_{\infty}^{-1},
\end{equation}
i.e. the surface potential is again controlled solely by $\lambda_D/\ell$.

\begin{figure}[h]
\begin{center}
\includegraphics[width=1\columnwidth]{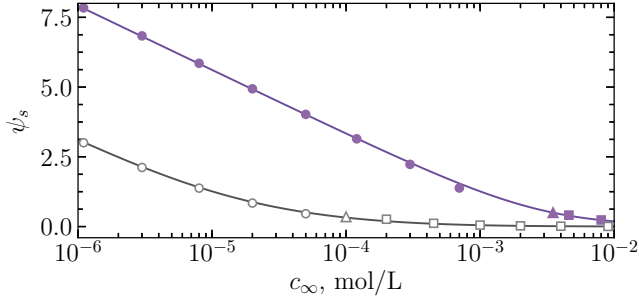}
\end{center}
\caption{$\psi_{s}$ vs $c_{\infty}$ computed using $H=15$ nm and $\ell = 5$ (upper solid curve) and 30 nm (lower solid curve).
Filled and open circles are obtained from Eqs.\eqref{eq:psi_s_large_rho_hh_dimensional} and \eqref{eq:psi_s_length}. Squares show predictions of Eq.~\eqref{eq:psi_s_ohshima}, and triangles mark the points of the curves, where $\lambda_D = \ell$.}\label{fig:Fig9}
\end{figure}

In Fig.~\ref{fig:Fig9} we plot $\psi_s$ vs. $c_{\infty}$ for these two specimen examples of the films. The surface potential is quite high at $c_{\infty} \simeq 10^{-6}$ mol/L (ca. 198 and 78 mV) and reduces with salt. At larger concentrations $\psi_s$ becomes smaller than unity and practically vanishes when  $c_{\infty} \geq 10^{-2}$ mol/L. To specify better the branches of low and high concentrations, in Fig.~\ref{fig:Fig9} we have marked $c_{\infty}^{\blacktriangle}$ and $c_{\infty}^{\vartriangle}$ by black and open triangles. 
For an upper curve computed using $\ell = 5$ nm this is located at $c_{\infty}^{\blacktriangle} \simeq 3.7 \times 10^{-3}$ mol/L, and for a lower, of $\ell = 30$ nm, at  $c_{\infty}^{\vartriangle} \simeq  10^{-4}$ mol/L. The corresponding surface potentials are $\psi_s \simeq 0.5$ and $0.3$. Thus, when $\lambda_D/\ell = 1$, both films are of low surface potentials. The first film is quasi-thick as discussed above, and the branch of the curve with $c_{\infty} \geq c_{\infty}^{\blacktriangle}$  is well fitted by Eq.\eqref{eq:psi_s_ohshima}. We recall that at $c_{\infty}^{\vartriangle}$ the second film still remains quasi-thin  and becomes quasi-thick, where \eqref{eq:psi_s_ohshima} should be strictly valid, only when $c_{\infty} \simeq 4.3 \times 10^{-4}$ mol/L (see Fig.~\ref{fig:Fig8}).
We see, however, that the fit is
quite good for $c_{\infty} \geq c_{\infty}^{\vartriangle}$, although at concentrations smaller than $4.3 \times 10^{-4}$ mol/L there is
some discrepancy, and Eq.\eqref{eq:psi_s_ohshima} slightly overestimates $\psi_s$. Also included in Fig.~\ref{fig:Fig9} are theoretical calculations for low salt concentrations. We see that at  $c_{\infty}$ smaller than $c_{\infty}^{\blacktriangle}$  Eq.\eqref{eq:psi_s_large_rho_hh_dimensional} is very accurate for a curve of $\ell = 5$ nm. When $\ell = 30$ nm, Eq.\eqref{eq:psi_s_length} provides an excellent fit to numerical data.
Finally, we would like to stress that it is impossible to generate a very high $\psi_s$ just by increasing $\varrho$. This is well seen in Fig.~\ref{fig:Fig9}, where the upper curve corresponds to the film with  36 times larger $\varrho$ than that for a film corresponding to a lower curve. The ratio of the values surface potentials  for these two coatings is always smaller. Its largest value is equal to 18, as follows from Eq.\eqref{eq:psi_s_ohshima} for the high salt regime, where $\psi_s$ is small. However, when $\psi_s$ is large, its amplification with $\varrho$ is very weak (only about 2 when $c_{\infty} \simeq 10^{-6}$ mol/L).

\begin{figure}[h]
\begin{center}
\includegraphics[width=1\columnwidth]{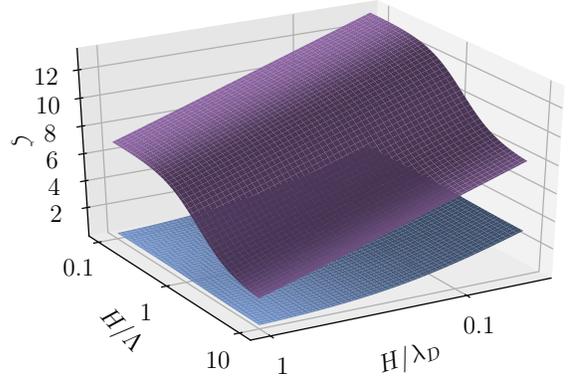}
\end{center}
\caption{$\zeta$ plotted as a function of two variables, $H/\lambda_D$ and $H/\Lambda$,  for a coating of thickness $H = 15$ nm using $\ell = 5$ (upper surface) and $\ell = 30$ nm (bottom surface). In the latter case $\zeta \simeq \psi_s$ as discussed in the text.}\label{fig:Fig10}
\end{figure}

We are now on a position to calculate $\zeta$, which generally depends on the Brinkman length $\Lambda$, and to contrast $\zeta$ to $\psi_s$. In Fig.~\ref{fig:Fig10} we plot $\zeta$ as a function of two variables, $H/\lambda_D$ and $H/\Lambda$, for two porous coatings discussed above. We recall that they are of the same thickness, but their values of $\ell$ are different. An overall conclusion from this three dimensional plot is that for a film of $\ell = 5$ nm the zeta potential is larger and very sensitive to $H/\Lambda$. However, for a coating of $\ell = 30$ nm the effect of $H/\Lambda$ on  $\zeta$, if any, is not discernible at the scale of Fig.~\ref{fig:Fig10}. Indeed, as discussed in Sec.~\ref{sec:small_rho_kH2}, in this case even at the infinite Brinkman length the zeta potential exceeds $\psi_s$, but very slightly (see also the lower curve in Fig.~\ref{fig:Fig6}). Simple calculations show that $\zeta - \psi_s \leq (H/\ell)^2$, which is equal to 0.25, i.e. very small, when $\ell = 5$ nm. In other words, for this film $\zeta \simeq \psi_s$ and can be evaluated, depending on $c_{\infty}$ that tunes an electrostatic regime, either using Eq.\eqref{eq:psi_s_length} or Eq.\eqref{eq:psi_s_ohshima}. Consequently, below we focus only on the quasi-thick film of $\ell = 5$ nm.

\begin{figure}[h]
\begin{center}
\includegraphics[width=1\columnwidth]{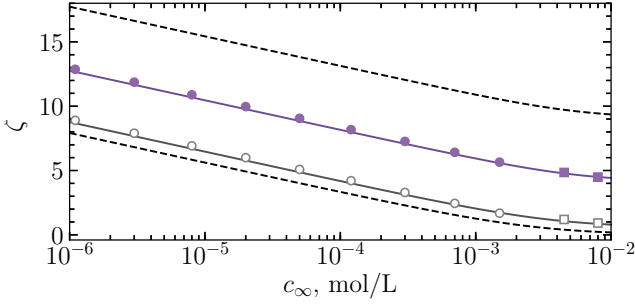}
\end{center}
\caption{$\zeta$ vs $c_{\infty}$ computed using $\Lambda = 30$ nm (upper solid curve) and 3.75 nm (lower solid curve) for a film of $H=15$ nm and $\ell = 5$ nm.
Filled and open circles are obtained from Eqs.~\eqref{eq:zeta_large_BL_large_rho_large_rho_hh_dimensional} and \eqref{eq:zeta_4}. Filled and open squares show calculations from Eqs.\eqref{eq:zeta_lin_smallKH} and \eqref{eq:zeta_lin_largeKH}.  Dashed curves show upper and lower bounds on $\zeta$.}
\label{fig:Fig11}
\end{figure}

We now compute the salt dependence of $\zeta$ for a film of $\ell = 5$ nm by setting $\Lambda = 30$ and 3.75 nm. These give $H/\Lambda = 0.5$ and 4, which should correspond to large and small Brinkman length regimes (see Fig.~\ref{fig:Fig6}).
The results of numerical calculations are shown in Fig.~\ref{fig:Fig11} together with the computed bounds on $\zeta$. We see that the difference between the upper
and lower bounds is quite large. The numerical $\zeta$-curves at finite $\Lambda$ are confined between these bounds, and are of the same shape as $\psi_s$, but shifted towards higher values that grow with  $\Lambda$ until $\zeta$ reaches its upper attainable limit.
When $c_{\infty} \leq c_{\infty}^{\blacktriangle}$ the surface potential $\psi_s$ is given by \eqref{eq:psi_s_large_rho_hh_dimensional}. At small $H/\Lambda$ the expression for the zeta potential can be
obtained  from Eq.~\eqref{eq:zeta_large_BL_large_rho_large_rho_hh_modified}
\begin{equation}\label{eq:zeta_large_BL_large_rho_large_rho_hh_dimensional}
\zeta \simeq \psi_s  + \frac{1}{2}\left( \dfrac{H}{\ell}\right)^2 \left[1 - \frac{5}{12} \left(\frac{H}{\Lambda}\right)^2\right] + 1,
\end{equation}
and when $H/\Lambda$ is large, it follows from \eqref{eq:zeta_small_BL_large_rho_large_rho_hh_modified} that
\begin{equation}\label{eq:zeta_4}
\zeta \simeq \psi_s  + \left( \dfrac{\Lambda}{\ell}\right)^2  + \dfrac{2 \Lambda}{H} 
\end{equation}
In Sec.~\ref{sec:general} we have clarified that the hydrodynamic permeability of the porous films $\propto H^2$ at low $H/\Lambda$ and $\propto\Lambda^2$ when $H/\Lambda$.  Thus, Eqs.\eqref{eq:zeta_large_BL_large_rho_large_rho_hh_dimensional} and \eqref{eq:zeta_4} point strongly that the ratio of the hydrodynamic permeability to $\ell^2$ is an important parameter controlling $\zeta$.
At larger concentrations, $c_{\infty} \geq c_{\infty}^{\blacktriangle}$, small $\psi_s$ is given by Eq.\eqref{eq:psi_s_ohshima}. Using then  Eqs.\eqref{eq:zeta_large_BL_oiv} and \eqref{eq:zeta_small_BL_next} for small and large $H/\Lambda$ we derive
\begin{equation}\label{eq:zeta_lin_smallKH}
  \zeta \simeq  \psi_s \left[ 2 - \frac{1}{4} \left(\frac{H}{\Lambda}\right)^2 \right] + \frac{1}{2} \left(\frac{H}{\ell}\right)^2\left[ 1- \frac{5}{12} \left(\frac{H}{\Lambda}\right)^2\right],
\end{equation}
\begin{equation}\label{eq:zeta_lin_largeKH}
  \zeta \simeq \psi_s \left(1 + \frac{2 \Lambda}{H} \right) + \left(\frac{\Lambda}{\ell}\right)^2
\end{equation}
We remark that again the ratio of the hydrodynamic permeability to $\ell^2$ becomes an important factor that determines the amplification of $\zeta$ compared to $\psi_s$.
The calculations from Eqs.\eqref{eq:zeta_large_BL_large_rho_large_rho_hh_dimensional}-\eqref{eq:zeta_lin_largeKH} are also included in Fig.~\ref{fig:Fig9} and we see that  provide an excellent fit to numerical data.

Thus, for quasi-thick films of a finite hydrodynamic permeability $\zeta \neq \psi_s$. As follows from Eqs.\eqref{eq:zeta_large_BL_large_rho_large_rho_hh_dimensional} - \eqref{eq:zeta_lin_largeKH}, besides $\psi_s$ (that can be tuned by varying the concentration of salt) the value of $\zeta$  also reflects $H/\Lambda$ and depends on the ratio of the hydrodynamic permeability to $\ell^2$.

\section{Concluding remarks}\label{sec:conclusion}

We have presented a theory of surface and zeta potentials of non-thick porous coatings, i.e. those of a thickness $H$ comparable
or smaller than that of the inner diffuse layer, of a finite hydrodynamic permeability. Our mean-field theory led to a number of  asymptotic approximations, which are both simple and very accurate, and can easily be used to predict or to interpret $\psi_s$ and $\zeta$ in different regimes, including situations when non-linear electrostatic effects become  significant.

The main results of our work can be summarized as follows. We have introduced an electrostatic length scale $\ell \propto \varrho^{-1/2}$ and
 demonstrated that depending on its value two different scenarios occur. In the high salt concentration regimes, $\ell > \lambda_D \propto c_{\infty}^{-1/2}$, the non-thick porous films are weakly charged and their electrostatic properties can be described by linearized equations. These films effectively behave either as thin or thick depending on the values of $H/\lambda_D$ and $H/\ell$. 
  We have also stressed the connection between the zeta potential and the Brinkman length, which is a characteristics of the hydrodynamic permeability of the porous film. Interestingly, the Brinkman length contribution to $\zeta$ permits to augment it compared to $\psi_s$ only if $(H/\ell)^2$ is large, i.e. when films are  quasi-thick.

Overall we conclude that tuning fluid transport inside a nanometric non-thick coating can dramatically affect the whole response of the large system to an applied electric field. Such a tuning can be achieved modifying its internal structure and charge density, or  by varying film thickness, or concentration of an external salt solution.

\begin{acknowledgments}

This work was supported by the Ministry of Science and Higher Education of the Russian Federation and by the German Research Foundation (grant 243/4-2) within the Priority Programme ``Microswimmers - From Single Particle Motion to Collective Behaviour'' (SPP 1726).
\end{acknowledgments}

\section*{Author's contribution}

 E.F.S. developed numerical codes, performed computations, and prepared the figures. N.B. participated in theoretical calculations. O.I.V. designed and supervised the project, developed the theory,  and wrote the
manuscript.

\section*{Data availability statement}
The data that support the findings of this study are available
within the article.

\appendix

\section{The limit of low potentials}\label{Ap:linear}

For completeness, in this Appendix we briefly discuss the case of low electrostatic potentials($\psi \leq $1), when Eq.~\eqref{eq:PB_io} can be linearized to give
\begin{equation}\label{eq:phi_in_LP}
\psi_{i, o}^{\prime \prime }=\kappa^2\left[\psi_{i, o} - \rho \Theta \left(H- z  \right)\right]
\end{equation}
Note that this case has been considered before by \citet{ohshima.h:1985}. Here we present a compact derivation of expressions for $\psi_0$ and $\psi_s$ in our (different) variables and complete the consideration by giving approximate expressions for $\Delta \psi$, $\mathcal{F}$. We also obtain an upper limit of $\zeta$.

Integrating Eq.~\eqref{eq:phi_in_LP} with prescribed boundary conditions (see Sec.~\ref{sec:general}) one can easily obtain

\begin{equation}\label{eq:phi_in_out_LP1}
\psi_{i}=\cosh(\kappa z)\left(\psi_0-\rho\right)+\rho, \,
\psi_{o}=\psi_s e^{-\kappa (z-H)},
\end{equation}
which leads to
\begin{equation}\label{eq:psi_0_LP}
\psi_0=\rho \left(1-e^{-\kappa H}\right)
\end{equation}

\begin{equation}\label{eq:psi_s_LP}
\psi_s=\rho \sinh(\kappa H) e^{-\kappa H}
\end{equation}

It is then straightforward to obtain
\begin{equation}\label{eq:delta_LP}
  \Delta \psi = \frac{\rho (1-e^{-\kappa H})^2}{2},
  \end{equation}
   and
\begin{equation}\label{eq:F_LP}
  \mathcal{F} = 1-e^{-\kappa H}
\end{equation}

We recall that these equations are valid at any $\kappa H$. At small $\kappa H$ they transform to Eq.\eqref{eq:small_rho_kappa_H}, but at large $\kappa H$ they reduce to
\begin{equation}\label{eq:ohshima}
 \psi_0 \simeq \rho, \, \psi_s \simeq \rho/2
\end{equation}
It is easy to verify that in fact the first equation of \eqref{eq:ohshima} is valid already when $\kappa H \geq 5$, and the second when $\kappa H \geq 2$. In other words, they describes not only the thick films, but also valid for some non-thick ones that can be termed quasi-thick.

Interestingly, low potential films satisfying \eqref{eq:ohshima} can potentially generate a high zeta-potential. Its upper achievable limit can be obtained using  Eq.\eqref{eq:v_s_large_BL} and is given by
\begin{equation}\label{eq:zeta_LP}
\zeta  \simeq \rho + \frac{\rho(\kappa H)^2}{2},
\end{equation}
The last equation coincides with that for weakly charged thick films~\cite{vinogradova.oi:2020}.
 Dividing \eqref{eq:zeta_LP} by  \eqref{eq:psi_s_LP} we conclude that for low potential thick and quasi-thick films $\zeta/\psi_s \simeq 2 + (\kappa H)^2$.



\begin{thebibliography}{47}
\bibitem [{not({\natexlab{a}})}]{note2}%
  \BibitemOpen
\bibinfo {note} {As a side note, since we
  consider here flat, infinite, and homogeneously charged porous layer, the
  electric field is the same everywhere (although the current density inside
  the porous layer is different from the outside) and normal flow is
  prohibited. In this case, one can neglect surface conduction (which tends to
  reduce electro-osmotic flow in finite size, curved, or heterogeneous systems)
  compared with bulk conduction ($Du = 0$), so we only discuss
  $v_{i,o}$.}\BibitemShut {Stop}%
\bibitem [{not({\natexlab{b}})}]{note3}%
  \BibitemOpen
\bibinfo {note} {This curve corresponds
  to a situation when the inner screening length is equal to $H$. So, formally
  it can also be seen as locus of a cross over between regimes of thick and
  thin films specified in Table~\ref{tbl:Ethickness}. However, since in our
  paper we focus only on the so-called non-thick films, we refer this curve to
  as a border between quasi-thin and quasi-thick coatings (see
  Sec.\ref{sec:theory_large_rho} for detail), which is more appropriate in our
  context.}\BibitemShut {Stop}%
\end{thebibliography}
%

\end{document}